\documentclass{ws-mpla}

\usepackage[centertags]{amsmath}
\usepackage{amsfonts}
\usepackage{amssymb}
\usepackage{latexsym}
\usepackage{textcomp}
\usepackage[superscript]{cite}

\RequirePackage[unicode,pdfstartview=FitH,colorlinks,bookmarksnumbered,%
bookmarksopen,bookmarksopenlevel=3]{hyperref}

\renewcommand{\vec}[1]{\ensuremath{\boldsymbol{\mathrm{#1}}}}

\begin{document}

\markboth{E.~B.~Manoukian \& N.~Yongram} {Polarization
Correlations in Pair Production from Charged and Neutral Strings}

%
\makeatletter
\def\catchline#1#2#3{\expandafter\def\expandafter\@clinebuf\expandafter
    {\@clinebuf\catchlinefont
    \noindent Modern Physics Letters A \hfill [\ArXivNo]\par
    \noindent Vol.~\textbf{20}, No.~8 (2005) 623--628\par
    \noindent \textcopyleft\ Authors 
    \hfill
    \href{http://dx.doi.org/10.1142/S0217732305015847}%
    {doi:10.1142/S0217732305015847}}\relax\par
    }%
\makeatother
\catchline{}{}{}{}{}
%

\newcommand{\ArXivNo}{\href{http://lanl.arxiv.org/abs/hep-th/0504195}{hep-th/0504195}}

\newcommand{\XXXSize}{{\fontsize{12}{12}\selectfont\fbox{\textbf{\ArXivNo}}}}
\newcommand{\XXXTitle}{\hfill\XXXSize\newline\vskip 0.4cm}

%
\def\draftnote{\today\quad\currenttime\quad [\ArXivNo]\qquad}%
%

\title{\MakeUppercase{Polarization Correlations in Pair Production from
Charged and Neutral Strings}}

\author{\footnotesize \MakeUppercase{E.~B.~Manoukian} and
\MakeUppercase{N.~Yongram}}

\address{School of Physics, \ Suranaree University of Technology,\\
Nakhon Ratchasima, 30000, Thailand\\
\email{\texttt{edouard@ccs.sut.ac.th}}}

\maketitle

\pub{Received 12 January 2004}{{}}

\begin{abstract}
Polarization correlations of $e^{+}e^{-}$ pair productions from
charged and neutral Nambu strings are investigated, via photon and
graviton emissions, respectively and explicit expressions for
their corresponding probabilities are derived and found to be
\textit{speed} dependent\@. The strings are taken to be circularly
oscillating closed strings, as perhaps the simplest solution of
the Nambu action\@. In the extreme relativistic case, these
probabilities coincide, but, in general, are different, and such
inquiries, in principle, indicate whether the string is charged or
uncharged\@. It is remarkable that these dynamical relativistic
quantum field theory calculations lead to a clear violation of
Local Hidden Variables theories\@.

\end{abstract}

\ccode{PACS Nos.: 11.25.Db,  12.20.Ds, 03.65.Ud, 13.88.+e}

\section*{{}}

We investigate the polarizations correlations of $e^{+}e^{-}$ pair
productions from charged and neutral Nambu strings via processes
of photon and graviton emissions, respectively\@. We consider
circularly oscillating closed strings as a cylindrical symmetric
solution\cite{B01,B02,B03,B04,B05,B06,B07,B08} arising from the
Nambu action\cite{B04,B05,B06,B07,B08,B09,B10,B11,B12}, as perhaps
the simplest string structures in field theory studies\@. Explicit
expressions are derived for their corresponding correlations
probabilities and are found to be \textit{speed} dependent\@. In
particular, due to the difference of these probabilities, in
general, inquiries about such correlations, would indicate whether
the string is charged or uncharged\@. In the extreme relativistic
case, however, these probabilities are shown to coincide\@. The
study of such polarization correlations are carried out in the
spirit of classic experiments, cf.~Refs.~\refcite{B13},
\refcite{B14}, to discriminate against Local Hidden Variable (LHV)
theories\@. In this respect alone, it is remarkable that our
explicit expressions of polarizations correlations, as obtained
from dynamical relativistic quantum field theories, are found to
be in clear violation with LHV theories\@. The speed dependence of
polarizations correlations is a common feature of dynamical
computations in quantum field theory\cite{B15,B16}\@.\\

The trajectory of the closed string is described by a vector
function $\vec{R}(\sigma,t)$, where $\sigma$ parameterizes the
string, satisfying\cite{B04,B05,B06,B07,B08,B09,B10,B11,B12}
\begin{align}
\ddot{\vec{R}}-\vec{R}''=0,\quad
\dot{\vec{R}}\cdot\vec{R}'&=0,\quad
\dot{\vec{R}}^{2}+\vec{R}'^{2}=1,\label{Eqn1}\\[0.5\baselineskip]
\vec{R}\left(\sigma+\dfrac{2\pi}{m},t\right)&=\vec{R}(\sigma,t),
\label{Eqn2}
\end{align}
where the mass scale $m$ is taken to be the mass of the electron,
$\dot{\vec{R}}=\partial\vec{R}/\partial t$,
$\vec{R}'=\partial\vec{R}/\partial\sigma$, with general solution
\begin{align}\label{Eqn3}
\vec{R}(\sigma,t)=\dfrac{1}{2}\left[\vec{A}(\sigma-t)+\vec{B}(\sigma+t)\big.\right],
\quad\vec{A}'^{2}=\vec{B}'^{2}=1.
\end{align} \\

We consider a solution of the
form\cite{B02,B03,B04,B05,B06,B07,B08}
\begin{align}\label{Eqn4}
\vec{R}(\sigma,t)=\dfrac{1}{m}\left(\cos m\sigma,\sin m\sigma,
0\big.\right)\sin mt,
\end{align}
with the $z$-axis perpendicular to the plane of oscillations\@.
For a string of total charge $Q$, this generates a current
density\cite{B05} $J^{\mu}(x)$  with structure
$\left(x=\left(t,\vec{r},z\right)\right)$
\begin{align}
J^{\mu}(x)&=\int\!\!\dfrac{\mathrm{d}^{2}\vec{p}}{(2\pi)^{2}}
\int^{\infty}_{-\infty}\!\dfrac{\mathrm{d}q}{(2\pi)}
\int^{\infty}_{-\infty}\!\dfrac{\mathrm{d}P^{0}}{(2\pi)}\;
\mathrm{e}^{\mathrm{i}\vec{p}\cdot\vec{r}}\mathrm{e}^{\mathrm{i}qz}
\mathrm{e}^{-\mathrm{i}P^{0}t}J^{\mu}(P^{0},\vec{p}),\label{Eqn5}\\[0.5\baselineskip]
J^{\mu}(P^{0},\vec{p})&=2\pi\sum_{N}\delta(P^{0}-mN)B^{\mu}(\vec{p},N),\label{Eqn6}
\end{align}
summing over integers,
\begin{align}
B^{0}(\vec{p},N)&=a_{N}J^{2}_{N/2}\left(\dfrac{|\vec{p}|}{2m}\right),
\label{Eqn7}\\[0.5\baselineskip]
\vec{B}(\vec{p},N)&=\dfrac{mN}{|\vec{p}|^{2}}\vec{p}B^{0}(\vec{p},N),
\label{Eqn8}\\[0.5\baselineskip]
a_{N}&=Q(-1)^{N/2}\cos\left(\dfrac{N\pi}{2}\right),
\label{Eqn9}
\end{align}
where $J_{N/2}$ are the ordinary Bessel functions of order
$N/2$\@.\\

We consider the process of $e^{+}e^{-}$ pair production via a
photon emission, given by the amplitude\cite{B04,B05,B06,B07,B08},
cf.~Refs.~\refcite{B17}, \refcite{B18}, up to an overall
multiplicative factor irrelevant for the problem at hand\@.
\begin{align}\label{Eqn10}
\mathcal{A}\propto
J^{\mu}(2p^{0},\vec{p}_{1}+\vec{p}_{2})\dfrac{1}{(p_{1}+p_{2})^{2}}
\left[\overline{u}(\vec{p}_{1},\sigma_{1})\gamma_{\mu}
v\vec{p}_{2},\sigma_{2})\big.\right]
\end{align}
with the four momenta of $e^{-}$, $e^{+}$, respectively, given by
\begin{align}
\vec{p}_{1}&=k(0,1,0),\quad \vec{p}_{2}=k(1,0,0),\quad
k=m\gamma\beta,
\label{Eqn11}\\[0.5\baselineskip]
p^{0}_{1}&=p^{0}_{2}=(\vec{k}+m^{2})^{1/2}\equiv p^{0}=m\gamma,
\label{Eqn12}
\end{align}
where $\gamma=1/\sqrt{1-\beta^{2}}$ is the Lorentz factor\@. The
measurement of the spin projection of the electron is taken along
an axis making an angle $\chi_{1}$ with the $z$-axis and lying in
a plane parallel to the $x$--$z$ plane,
\begin{align}\label{Eqn13}
u=\sqrt{\dfrac{p^0+m}{2m}}\begin{pmatrix}
\xi_{1}\\\\
\dfrac{k\sigma_{2}}{p^0+m}\xi_{1}
\end{pmatrix},\qquad v=\sqrt{\dfrac{p^0+m}{2m}}\begin{pmatrix}
-\dfrac{k\sigma_{1}}{p^0+m}\xi_{2}\\\\ \xi_{2}
\end{pmatrix},
\end{align}
where the direction of the spin of the positron lies in a plane
parallel to the $y$--$z$ plane\@. For the $2$-spinors, we have
\begin{align}\label{Eqn14}
\xi_{1}=\begin{pmatrix}
\cos\left(\dfrac{\chi_{1}}{2}\right)\\\\
-\sin\left(\dfrac{\chi_{1}}{2}\right)
\end{pmatrix},\qquad \xi_{2}=\begin{pmatrix}
\sin\left(\dfrac{\chi_{2}}{2}\right)\\\\
\cos\left(\dfrac{\chi_{2}}{2}\right)
\end{pmatrix}.
\end{align} \\

A tedious but straightforward computation gives
\begin{align}\label{Eqn15}
\mathcal{A}\propto\dfrac{\left[-\mathrm{i}\left(1-\frac{\gamma\beta^{2}}{2}\right)
\cos\left(\frac{\chi_{1}-\chi_{2}}{2}\right)+\left(1+\frac{\gamma\beta^{2}}{2}\right)
\cos\left(\frac{\chi_{1}+\chi_{2}}{2}\right)\right]}{\left(\gamma^{2}\beta^{2}-2\right)}
\sum_{N}\delta(2p^{0}-mN)B^{0},
\end{align}
where we note that $2p^{0}/m=2\gamma$ is quantized\@.
\textit{Given} that the above process has occurred, a standard
computation, as in~Refs.~\refcite{B15}, \refcite{B16}, given the
following explicit expression for the probability of the
simultaneous measurements of the spins of $e^{-}$, $e^{+}$, with
angles $\chi_{1}$, $\chi_{2}$, as specified above,
\begin{align}\label{Eqn16}
P[\chi_{1},\chi_{2}]=\dfrac{\left(2\sqrt{1-\beta^{2}}-\beta^{2}\right)^{2}
\cos^{2}\left(\frac{\chi_{1}-\chi_{2}}{2}\right)+
\left(2\sqrt{1-\beta^{2}}+\beta^{2}\right)^{2}
\cos^{2}\left(\frac{\chi_{1}+\chi_{2}}{2}\right)}{4\left(2-\beta^{2}\right)^{2}}
\end{align}
the so-called probability of the polarizations correlations of the
emitted pair and is \textit{speed} dependent\@. If the spin of
only one of the particles, say, that of $e^{-}$, is measured, then
we have to sum (\ref{Eqn16}) over the two possible outcomes for
$e^{+}$: $\chi_{2}$, $\chi_{2}+\pi$, for a given $\chi_{2}$, i.e.,
for the probability of measuring the spin of $e^{-}$ only, we have
\begin{align}\label{Eqn17}
P[\chi_{1},-]=P[\chi_{1},\chi_{2}]+P[\chi_{1},\chi_{2}+\pi]=\frac{1}{2}.
\end{align}
Similarly, for the probability $P[-,\chi_{2}]$, where only a
measurement of the spin of $e^{+}$ is made, we obtain
\begin{align}\label{Eqn18}
P[-,\chi_{2}]=\frac{1}{2}.
\end{align} \\

In the extreme relativistic case $\beta\rightarrow 1$,
(\ref{Eqn16}) gives
\begin{align}\label{Eqn19}
P[\chi_{1},\chi_{2}]\longrightarrow\dfrac{1}{4}\left[
\cos^{2}\left(\dfrac{\chi_{1}-\chi_{2}}{2}\right)+
\cos^{2}\left(\dfrac{\chi_{1}+\chi_{2}}{2}\right)\right].
\end{align} \\

The neutral string, of a given mass $M$, generates an
energy-momentum tensor density $T^{\mu\nu}(x)$ with structure
\cite{B06}
\begin{align}
T^{\mu\nu}(x)&=\int\!\!\dfrac{\mathrm{d}^{2}\vec{p}}{(2\pi)^{2}}
\int^{\infty}_{-\infty}\!\dfrac{\mathrm{d}q}{(2\pi)}
\int^{\infty}_{-\infty}\!\dfrac{\mathrm{d}P^{0}}{(2\pi)}\;
\mathrm{e}^{\mathrm{i}\vec{p}\cdot\vec{r}}\mathrm{e}^{\mathrm{i}qz}
\mathrm{e}^{-\mathrm{i}P^{0}t}T^{\mu\nu}(P^{0},\vec{p}),
\label{Eqn20}\\[0.5\baselineskip]
T^{\mu\nu}(P^{0},\vec{p})&=2\pi\sum^{\infty}_{N=-\infty}
\delta(2p^0-mN)B^{\mu\nu}(\vec{p},N),
\label{Eqn21}\\[0.5\baselineskip]
B^{00}(\vec{p},N)&= \beta_N J^{2}_{N/2}(z),\quad
z=\dfrac{|\vec{p}|}{2m},
\label{Eqn22}\\[0.5\baselineskip]
B^{0a}(\vec{p},N)&= \beta_N \dfrac{P^{0}p^{a}}{|\vec{p}|^{2}}
J^{2}_{N/2}(z),\quad a=1,2
\label{Eqn23}\\[0.5\baselineskip]
B^{ab}(\vec{p},N)&= \beta_N\left(A_N\delta^{ab}+E_N\dfrac{p^{a}
p^{b}}{|\vec{p}|^{2}}\right),\quad a,b=1,2
\label{Eqn24}\\[0.5\baselineskip]
B^{\mu3}(\vec{p},N)&= 0,\quad \mu=0,1,2,3
\label{Eqn25}\\[0.5\baselineskip]
A_N &=
\dfrac{1}{4}\left[J_{\frac{N}{2}+1}(z)-J_{\frac{N}{2}-1}(z)\right]^{2},
\label{Eqn26}\\[0.5\baselineskip]
E_N &=J_{\frac{N}{2}+1}(z)J_{\frac{N}{2}-1}(z),
\label{Eqn27}\\[0.5\baselineskip]
\beta_N &=
M(-1)^{N/2}\cos\left(\frac{N\pi}{2}\right).
\label{Eqn28}
\end{align} \\

For $e^{+}e^{-}$ pair production via the emission of a graviton,
the amplitude of the process is given by
\begin{align}\label{Eqn29}
\mathcal{A}\propto
T^{\sigma\lambda}(2p^{0},\vec{p}_{1}+\vec{p}_{2})
\dfrac{\left[g_{\sigma\mu}g_{\lambda\nu}
-\frac{1}{2}g_{\sigma\lambda}g_{\mu\nu}\right]}{\left(p_{1}+p_{2}\right)^{2}}
T^{\mu\nu}_{e^{+}e^{-}},
\end{align}
where $T^{\mu\nu}_{e^+e^-}$ is the energy-momentum tensor density
associated with the pair\@. From (\ref{Eqn21})--(\ref{Eqn22}),
(\ref{Eqn11}), (\ref{Eqn12}), this simplifies to
\begin{align}\label{Eqn30}
\mathcal{A}\propto\dfrac{1}{\left(p_{1}+p_{2}\right)^{2}}\left\{-2m\overline{u}vT^{00}+
2\left[\left(\overline{u}\gamma_{a}v\right)\left(p^{1}_{b}-p^{2}_{b}\right)+m\delta_{ab}
\overline{u}v\big.\right]T^{ab}\Big.\right\}.
\end{align} \\

The recurrence relation
\begin{align}\label{Eqn31}
J_{\frac{N}{2}-1}(z)=\frac{2\sqrt{2}}{\beta}J_{\frac{N}{2}}(z)-J_{\frac{N}{2}+1}(z),
\end{align}
allows one to express $A_{N}$, $E_{N}$ in terms of $J_{N/2}+1$,
$J_{N/2}$ which differ by one order only, and for sufficiently
high energies, they may be expressed in terms of $J_{N/2}$\@.\\

All told, given that the above process has occurred, a direct
straightforward simplification of the expression in (\ref{Eqn30})
leads to the following expression for the polarizations
correlations probability of the pair
\begin{align}\label{Eqn32}
P[\chi_{1},\chi_{2}]=\dfrac{1}{4\left(4-\beta^{2}-2\sqrt{2}\beta\right)}
&\Bigg[\left(\sqrt{2(1-\beta^{2})}-\sqrt{2}+\beta\right)^{2}
\cos^{2}\left(\dfrac{\chi_{1}-\chi_{2}}{2}\right)\nonumber\\[0.5\baselineskip]
&+\left(\sqrt{2(1-\beta^{2})}+\sqrt{2}-\beta\right)^{2}
\cos^{2}\left(\dfrac{\chi_{1}+\chi_{2}}{2}\right)\Bigg]
\end{align}
and again is speed dependent, $P[\chi_{1},-]=1/2=P[-,\chi_{2}]$
for a measurement of the spin of only one of the particles\@. The
fact that the polarizations correlations probabilities of the
$e^{+}e^{-}$ pair emitted from the charged and neutral strings are
different in general,  such inquiries indicate, in principle,
whether the string is charged or uncharged\@. In the extreme
relativistic case, the probability in (\ref{Eqn32}) coincides with
the expression on the right-hand side of (\ref{Eqn19}) for a
charged string\@.\\

Finally we note that these dynamical relativistic quantum field
theory calculations lead to a violation of LHV
theories\@.\cite{B13,B14} To this end, define:\cite{B13,B14}
\begin{align}\label{Eqn33}
S=P[\chi_1,\chi_2]-P[\chi_1,\chi'_2]+P[\chi'_1,\chi_2]+P[\chi'_1,\chi'_2]
-P[\chi'_1,-]-P[-,\chi_2]
\end{align}
for two pairs of angles $(\chi_1,\chi_2)$, $(\chi'_1,\chi'_2)$\@.
To show violation with LHV theories, it is sufficient to consider
one experimental situation which gives for $S$ a value
outside\cite{B13,B14} the interval $[-1,0]$\@. To this end for
$\beta=0.8$, $\chi_1=0^{\circ}$, $\chi_2=160^{\circ}$,
$\chi'_1=100^{\circ}$, $\chi'_2=10^{\circ}$, we obtain $S=-1.088$,
$S=-1.103$ for the charged and neutral strings, respectively,
leading to a clear violation of LHV theories\@.\\

\section*{Acknowledgments}

The authors acknowledge with thanks for being granted a ``Royal
Golden Jubilee Ph.D. Program'' by the Thailand Research Fund
(Grant No.~PHD/0022/2545) for especially carrying out this
project\@.

\end{document}